\documentclass[prl,twocolumn,showpacs,preprintnumbers,amsmath,amssymb]{revtex4}

\usepackage{graphicx}
\usepackage{dcolumn}
\usepackage{bm}

\usepackage{epsf}

\newcommand{\be}{\begin{equation}}
\newcommand{\ee}{\end{equation}}
\newcommand{\beqn}{\begin{eqnarray}}
\newcommand{\eeqn}{\end{eqnarray}}

\begin{document}

\title{Predictions of the residue cross-sections for the elements Z=113 and Z=114}

\author{ Bertrand Bouriquet} 
 \affiliation{Yukawa Institute for Theoretical Physics, Kyoto 
University,Kyoto 606-8502, Japan}
\author{ Yasuhisa Abe}
 \affiliation{Yukawa Institute for Theoretical Physics, Kyoto 
University,Kyoto 606-8502, Japan}
\author{Grigori Kosenko}
 \affiliation{ Department of Physics, University of Omsk, RU-644077 Omsk, Russia }
\date{Submitted to Phys. Rev. Let. on 4 August 2003 }

\begin{abstract}
An extremely good reproduction of experimental excitation function of the  $1n$ reactions producing  Z=110,Z=111 and Z=112 is obtained by the two-step model and the statistical decay code KEWPIE. Thus, an extension of the recipe permits us to predict reliable values of the residue cross-sections of the elements Z=113 and Z=114 which will be a useful guide for planning of experiments. 
\end{abstract}

\maketitle

\section {Introduction}
The production and the identification of new elements have been made since the discovery of the periodic table of the elements by Mendelejeff in 1869 \cite{Men}. Nowdays we are still looking for the heaviest elements, so called super-heavy elements(SHE). The properties and the conditions of the existence of the SHE have been foreseen since the establishment of the nuclear shell model \cite{May}. The studies of shell structure predict an island of stability for those elements around the double closed shell next to $^{208}Pb$. Several values of the magic number for Z have been predicted between 114 and 126, depending on the model used in the calculations, while that for N is 184 \cite{Smo,Sobi,Mol,Rei,Ben}. However, the major difficulty in the production of the SHE lies in the lack of knowledge about the reaction mechanism, especially about fusion mechanism of the massive systems. In the region of the heavy nuclei, the so-called fusion-hindrance is known to exist experimentally\cite{Qui,Sch,Res}. But no complete theoretical description of the mechanism is yet given. As a consequence, it has been difficult to predict reliably which combination of target, projectile and bombarding energy is optimum to produce new elements. 

Let's assume the compound nucleus theory\cite{Boh}, though it might not be fully justified for such unstable systems. We can express the residue  cross-section as

\be
\sigma_{res} = \pi {\bar{\lambda}}^2 \sum (2J+1) P^J_{fusion}(E_{c.m.}) . P^J_{surv}(E^*),  \label{fcs}
\ee

where $J$ denotes a total spin of the system and $E_{c.m.}$ the energy in the center of mass. $E^*$ is equal to $E_{c.m.} + Q$ with the fusion Q-value. As usual $P^J_{fusion}$ and $P^J_{surv}$ are the fusion and the survival probabilities for the spin $J$, respectively. The latter is  obtained by the use of the statistical disintegration code, which is newly developed by solving the time-dependent Bateman equation \cite{Bou}. The former is the most unknown part. But recently, one of the present authors {\it et al.} have proposed the two-step model, where the fusion process is divided into two phases: the approaching phase and the formation phase. They correspond respectively  to the passing over the Coulomb barrier and to the shape evolution toward  the spherical compound nucleus, starting from the pear-shaped sticking configuration of the incident system. Thus, 

\be
P^J_{fusion} = P^J_{sticking} . P^J_{formation} 
\ee  

The calculation of the probabilities can be done by solving dynamics of the process in each step. The amalgamated system is supposed to be exited, which mean the incident kinetic energy has been converted into the thermal energy. This conversion of kinetic energy would be done suddenly at the contact point or would start before the top of the Coulomb barrier. We will assume the latter hypothesis, as it is well known that the dissipation observed in the deep inelastic collision (DIC) is understood by the viewpoint \cite{Gro}. With this assumption we calculate the sticking probability with a Langevin equation including a frictional force and associated fluctuation with a time dependent temperature \cite{Fro}. The formation probability is calculated by a multi-dimensional Langevin equation for the shape evolution with a constant temperature. The full description of this model has been given in the references \cite{She,Abe2}.  

We will first make a brief theoretical description of the two-step model. Then, we will check the ability of the model to reproduce known experimental data. Finally we will present predictions for the residue cross-sections for the reaction $^{70}Zn + ^{209}Bi \rightarrow ^{278}113 + 1n $, $^{71}Ga + ^{208}Pb \rightarrow ^{278}113 + 1n $ and   $^{76}Ge + ^{208}Pb \rightarrow ^{283}114 + 1n $   that have not yet been measured but should be in a near future.   

\section{Reminder of the Two-step model for fusion}
We briefly recapitulate here the theoretical framework of the two-step model used to calculate the fusion probability.

\subsection{Approaching phase}
To  describe the approaching phase, a classical description of the relative motion between colliding nuclei is used with a frictional force and an associate fluctuation force. In the original treatment  \cite{Gro,Ban} the random force was not implemented, but is properly taken into account in the present model. 

\beqn
\left\{
\begin{array}{@{}cccc@{}}
\mu \frac{{d^2 r}}{dt^2} & = & -  \frac{\partial V}{\partial r} -  \frac{\partial }{\partial r} \frac{\hbar^2 L^2 }{2 \mu r^2} - C_r(r) \frac{dr}{dt} + \Gamma_r(t)  \\
\frac{dL}{dt}  & = & \frac{C_T(r)}{\mu} [ L-\frac{5}{7} L_0  ] + \Gamma_T(t)  
\end{array} \right. \\
<\Gamma_i(t) \Gamma_j(t)>  =  2 T(t).C_i(r(t)).\delta_{ij} . \delta(t-t'), \label{diflu}
\eeqn
   
where $\mu$ is the reduced mass and $V$ is the sum of Coulomb and nuclear attractive potential. $C_i(r)$ is the radial and the tangential friction, respectively. We will notice that here the rolling friction is neglected. $L_0$ is the incident angular momentum and $\frac{5}{7}L_0$ so-called sliding limit. $\Gamma_i (t)$ denotes a Gaussian random force with zero mean value. The equation \ref{diflu} is the dissipation-fluctuation theorem. In the case $i=j= \phi$, $r^2$ factor is necessary for $C_{\phi}(r)$ in {\it Eq.} \ref{diflu}. In this modelisation the temperature increases as the energy is dissipated by the friction force.

This modelisation is the one given by the surface fiction model (SFM)\cite{Gro}, which has been used successfully to reproduce the main features of the deep inelastic collisions,  expect the spin of the outgoing fragments which appear to require an inclusion of the rolling friction. Moreover the very different strengths between the radial and tangential frictional forces may not be physical, but may be a certain effective description. And of course, its applicability is limited to the energy region above the Coulomb barrier, due to its classical nature. Nevertheless, it is simple and has no free parameters, so we employ it for a description of the approaching phase for the realisation of the two-step model \cite{Kos}. 

One of the characteristic results of the model is the Gaussian distribution of the radial momentum at the touching point. The origin of this distribution comes from the Gaussian random force associated to the friction force. The width of this distribution is consistent with the temperature determined by the internal energy transferred from the incident kinetic energy. This distribution is used as an initial condition for the subsequent dynamics of the shape evolution from a pear shape toward the spherical shape. 

\subsection{Shape evolution}
The pear-shaped system formed by the contact of the nuclei involved in the collision is located outside of the conditional saddle point \cite{Swi,Abe6}. As a consequence, we have to solve the shape evolution toward the spherical shape under a friction force associated with a random force. The dynamics of this process is again treated by the general multidimensional Langevin equation \cite{Abe3,Abe4,Wad,Abe5}.

\beqn
\left\{
\begin{array}{@{}lll@{}}
 \frac{{dq_i}}{dt} & = & (m^{-1})_{ij} . p_j  \\
\frac{dp_i}{dt}  & = & -\frac{\partial V^J}{\partial q_i } -\frac{1}{2} \frac{\partial}{\partial q_i}    (m^{-1})_{jk} . p_j . p_k  \\ & & - \gamma_{ij} .  (m^{-1})_{jk} . p_k + g_{ij} . \Gamma_j(t)
\end{array} \right. \\
<g_{ik} g_{jk}>   =   \gamma_{ij} . T^J , \label{diflu2}
\eeqn

where $V^J$ denotes the liquid drop model (LDM) potential energy surface plus the rotational energy for spin $J$ \cite{Iwa,Sat}. The equation \ref{diflu2} is again the dissipation-fluctuation theorem with temperature $T ^J$ being assumed constant for the simplicity. The friction tensor is calculated by using the so-called one-body model (OBM) \cite{Blo}, that is, a one-body wall-and-window formula with  the two-center parametrisation of nuclear shape. 

  The ratio between the number of trajectories reaching the spherical configuration and the total number of trajectories gives the formation probability. Mostly, the trajectories do not go over the ridge line but go back to re-separation. We calculate this probability for various initial momenta at a given touching point. Then, the total formation probability is obtain by the convolution of those probabilities with the initial distribution of momenta obtained in the analyses of the approaching phase. 

\begin{figure}[htbp]
\epsfxsize=9cm
$$
\epsfbox{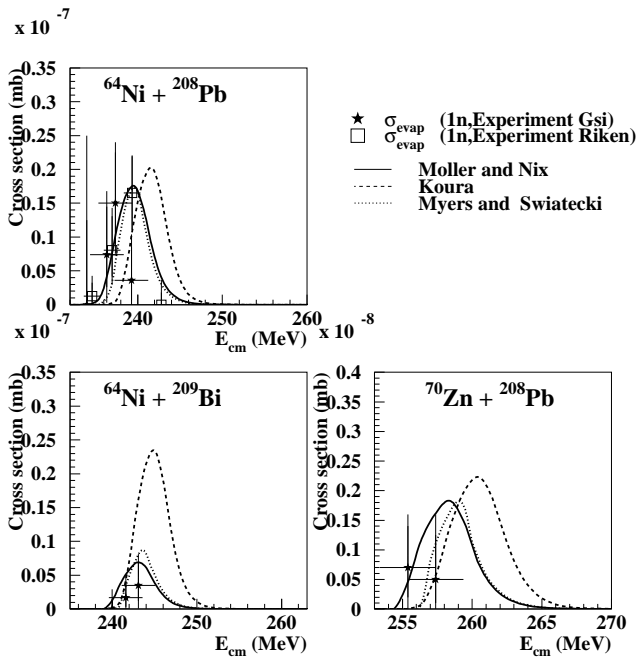}
$$
\caption{\it Residue cross-sections for systems that have been already studied experimentally. The symbol represent experimental data with the error bas in energy (thickness of the target) and cross section (sensitivity of experimental device and cumulated dose). The line represent the results of calculation assuming different table of mass and shell correction energy.
 \label{fig1} }
\end{figure}

\section{Comparison with experiment}

As the fusion probability is calculated as above,  we will now calculate residue cross-sections according to {\it Eq.} \ref{fcs}. In the scope of this paper we will only consider the $1n$ reactions. The decay calculations are done for each partial cross-section for a given $J$ by the code KEWPIE \cite{Bou} which has only one free intrinsic parameter that is the reduced friction coefficient ($\beta$) used for the Kramers factor \cite{Kra}. This value is set to  $\beta = 5. 10^{-21} s$  for all the results presented in this publication in consistence with OBM. Here, the essential ambiguities are the masses and shell correction energies employed for SHE. 
To know their influence  we have made the calculations employing three different tables of mass : Moller \& Nix, Koura, Myers \& Swiatecki \cite{Mol,Kou,Ms1}.  

First, we will take  $^{64}Ni +  ^{208}Pb   \rightarrow ^{271}110 + 1n $ system as an example for which more precise excitation function  is recently measured at RIKEN, in addition to the previous GSI data. Calculations are made with the fusion probabilities obtained above and the survival probabilities with KEWPIE code. The calculated peak positions in $E_{c.m.}$ are not far from the data, though there are small difference among the mass tables. This is mainly due to the slight difference in the macroscopic part of the masses among the tables, which is consistent with the qualitative explanation by Swiatecki {\it et al.} \cite{Swi2}.
Absolute values of the peaks of $1n$ cross-sections are larger than the experiment by a few to several orders of magnitude depending on the mass tables. Nevertheless, it would be remarkable that the calculation of the present model result in rather good reproductions of the experiments without using any free parameter adjusted. For more precise reproduction and thus accurate  predictions, however, we have to refine the model, in the calculation of $P_{sticking}$ and/or $P_{form}$, or to find accurate values of the masses etc. But it would be practical to introduce a phenomenological reduction factor, for example, a constant renormalization of the cross-sections. In view of a variety of predictions of the shell correction energy, we insteads introduce a phenomenological factor for the shell correction energy which is to be used all through each mass table employed, keeping its own characteristic tendency predicted over the elements and the isotopes under investigation. An extreme sensitivity of the calculated peak height to the shell correction is another reason for the factor effective in the systematic analyses.

 In order to reproduce the peak heights of the experimental measurements for the reaction $^{64}Ni +  ^{208}Pb   \rightarrow ^{271}110 + 1n $, the  factor is 0.4 , 0.36 and 0.4 for the shell correction energy of Moller \& Nix, Koura, Myers \& Swiatecki, respectively. 

 
In the Fig. \ref{fig1} is shown the comparison between the calculations and the experiments \cite{Hofm,Mori}. The  energy of the experimental data plotted in this figure corresponds to the energy in the center of mass minus the energy loss at the middle of the target. The associated error bars in energy correspond to the energy losses in the target.  

In the  upper left panels of the Fig. \ref{fig1} we notice that the position of the peak in $E_{c.m.}$ is well reproduced, especially for the tables of  Moller \& Nix and Myers \& Swiatecki. For a justification of the modification factor for of the shell correction we have made calculations for the systems  $^{64}Ni +  ^{209}Bi  \rightarrow ^{272}111 + 1n $ and $^{70}Zn +  ^{208}Pb  \rightarrow  ^{277}112 + 1n $,  keeping the same value of the multiplication factor for the shell correction energy. 
The lower panels of the figure \ref{fig1} show an overall (calculated curve within the experimental error bar) good agreement between the experimental measurements and the theory. Moreover, the global phenomenological  decreasing trend of the peak heights of the cross-sections is well reproduced, assuming the Moller \& Nix and Myers \& Swiatecki. For those two mass tables we can even say that a remarkably good agreement between the experiment and theory is obtained. We should remind that the same model has been already applied to the hot fusion path ($^{48}Ca$ + actinide reaction) and resulted in a fairly good reproduction of the data \cite{She,Abe2}.

\section{Predictions of the residue cross section for elements Z=113 and Z=114}{\label{pred}}

As we are able to reproduce very well the residue cross-section for the above known systems we can extend this recipe to the systems where measurements are not yet made.  We will focus on the yet  undiscovered element Z=113 and a new isotope of the Z = 114 produced by the cold fusion path. In order to calculate residue cross-sections we will just extend the calculations with the same factor for the shell correction energy as that we have set for the  $^{64}Ni +  ^{208}Pb  \rightarrow   ^{271}110 + 1n $ reaction.

\begin{figure}[htbp]
\epsfxsize=9cm
$$
\epsfbox{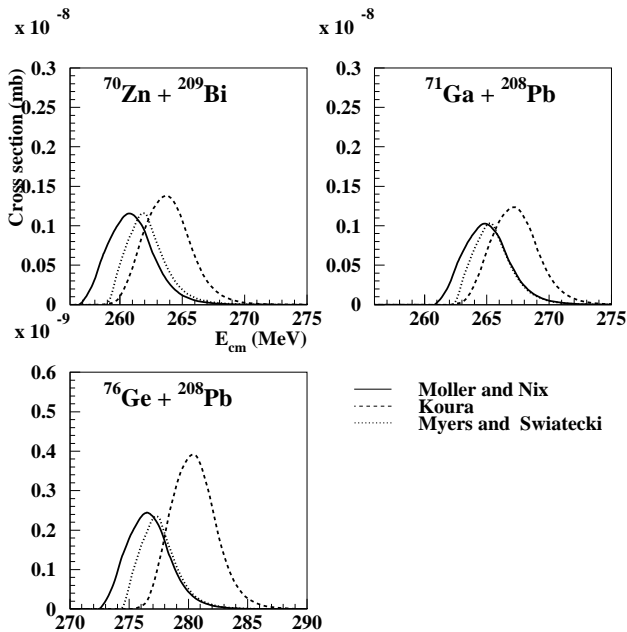}
$$
\caption{\it $1n$ Residue cross-section for system that have not been yet measured. The line represent the results of calculation assuming different table of mass and shell correction energy.
 \label{fig2} }
\end{figure}

For the production of the Z=113 the reaction  $^{70}Zn +  ^{209}Bi \rightarrow  ^{278}113 + 1n $  or the reaction  $^{71}Ga +  ^{208}Pb \rightarrow  ^{278}113 + 1n $ is under consideration. As we see on the upper panels of Fig. \ref{fig2}, both reactions are predicted to lead to roughly the same residue cross-sections for all the mass predictions employed. This similarity come from the fact that on one hand the  fusion barrier is unfavourable for the reaction  $^{71}Ga +  ^{208}Pb \rightarrow  ^{271}113 + 1n $  but on the other hand the high Q-value of this reaction leads to lower excitation energy that is favourable for the emission of neutron in the competition between fission and evaporation.

Moreover, the results are in agreement with the global phenomenological trend of a decreasing of the $1n$ residue cross-sections by roughly a factor $3$ for each increasing of one charge unit in the SHE.

The production of the element Z=114 is the other case that we will predict.  This element has been already produced by the hot fusion path \cite{Oga} but the experiment with the reaction $^{76}Ge +  ^{208}Pb  \rightarrow  ^{283}114 + 1n $  will be the first measurement of the production of $Z=114$ by the way of a $1n$ reaction (cold fusion path).  The results of the calculation are shown in  the lower panel of Fig. \ref{fig2}. The predicted value is around a few tenth of pico-barn. To measure the residue with a so low cross-section is a challenge in experiment.

\section{Conclusion}

As we are able to reproduce rather well both the positions and the absolute values of the peaks of measured $1n$ residue cross-sections by introducing only one free parameter, one reduction factor for the predicted shell correction energies, the two step model appears to be quite promising. Based on the success we have made the prediction on the $1n$ cross-sections in $^{70}Zn +  ^{209}Bi \rightarrow  ^{278}113 + 1n $, $^{71}Ga +  ^{208}Pb \rightarrow  ^{278}113 + 1n $ and  $^{76}Ge +  ^{208}Pb \rightarrow  ^{283}114 + 1n $ without adjusting anything more. This is the first predictions of the excitation function of residue cross section for Z=113 and 114 by the calculation with the dynamical model of reactions.
Of course, the ultimate test of the model is the future experiment performed on the above systems. The model is expected to provide a reliable guideline for future experiments for productions of super-heavy elements. Hopefully the present model permits  have an overview of the most suitable experiments and win a lot of experimental time.

\begin{acknowledgments}
The authors thank W.J.Swiatecki for fruitful discussions and K.Morita for the access to the data before publication. 
One of us (B.B) thanks the Yukawa Institute for its warm hospitality. 
The authors acknowledge JSPS for supports (contracts P-01741, 1340278). 
\end{acknowledgments}

\end{document}